\documentstyle[aps,prd]{revtex}
\tighten

\def\Lie{\hbox{\it\char'44}}

\def\be{\begin{equation}}
\def\ee{\end{equation}}
\def\ba{\begin{eqnarray}}
\def\ea{\end{eqnarray}}
\begin{document}
\draft

\title{A Linear-Nonlinear Formulation of Einstein Equations for the\\
Two-Body Problem in General Relativity}
\author{Pablo Laguna}

\address{
Department of Astronomy \& Astrophysics and \\
Center for Gravitational Physics \& Geometry\\
Penn State University, University Park, PA 16802, USA}

\date{\today}

\maketitle

\begin{abstract}
A formulation of Einstein equations is presented
that could yield advantages in the study of collisions of binary compact objects
during regimes between linear-nonlinear transitions.
The key idea behind this formulation is a separation of the dynamical variables 
into i) a fixed conformal 3-geometry,
ii) a conformal factor possessing nonlinear dynamics and iii)
transverse-traceless perturbations of the conformal 3-geometry.
\end{abstract}

In the two-body problem in General Relativity, involving black holes and/or neutron stars,
one can identify three main phases according to the extent to which nonlinear
effects play a fundamental role. When the two bodies are sufficiently far apart,
the binary system is in an {\it Inspiral} phase.
During this phase, a Post-Newtonian approach to gravity provides a suitable
framework. Loses due to gravitational radiation drive the binary system 
to the next stage, the {\it Merging} phase. During the merging,
which includes the last few orbits
and coalescence of the binary system,
nonlinear effects become essential and eventually dominate 
the dynamics of the system; thus,
the full set of Einstein fields equations is required.
The last stage of the problem is the {\it Ringdown} phase.
Here once again, the two-body problem can be treated perturbatively.
In this situation, however, the system can be investigated  as perturbations about
the outcome of the collision, namely a single black hole or neutron star.
This point of view is also known as the ``close-limit" and has been
shown to be a remarkable good approximation \cite{price-pullin}.

From the computational point of view, it is safe to say that the machinery, necessary
to study the {\it Inspiral} and {\it Ringdown} phases of the problem, is
available and reasonably well understood.
On the other hand, in spite of advances in computer technology and
numerical algorithms, the {\it Merging} phase of black hole and neutron star collisions
remains essentially unsolved;
that is, simulations that bridge the ``gap" between
the {\it Inspiral} and {\it Ringdown} stages are not yet available.
The expectation is that this situation will prevail for the next few years.
Furthermore, even if full nonlinear numerical simulations 
of the collision of compact objects are achieved \cite{Alliance}, they will most likely not be able to
cover, at least initially, the dynamical 
time-scales necessary to grab initial data from the {\it Inspiral} phase,
evolve it through the {\it Merging} phase and provide initial data for
{\it Ringdown} calculations. 
To get a sense of the formidable task involved, one only needs to keep in mind
that, for numerical approaches using asymptotically
inertial coordinates, the dynamical time-scale is $\sim M$.
This time-scale should be compared with $ \sim 1200\, M$,
the time-scale of numerical evolutions 
to cover the {\it Inspiral-Ringdown} gap.
Because of this foreseen limitation, it is important to 
develop methodologies that directly address the intermediate 
regimes between the main phases discussed above. 

The {\it Inspiral-Merging} intermediate regime has recently received significant attention.
In the context of black hole collisions, this is called the ``intermediate
binary black holes" problem by Thorne and collaborators \cite{Brady}.
They have proposed an approach 
in which numerical calculations are performed in co-rotating
coordinates with the binary system.
In this co-rotating frame, the metric evolves on long time-scales, namely
those of the inspiral scales. Furthermore, gauge conditions are chosen
such that gravitational degrees of freedom that are not excited by
radiation remain frozen.

Finding a reduced formulation of Einstein equations for
the {\it Inspiral-Merging} as well as {\it Merging-Ringdown}
intermediate regimes
is the focus of this paper, namely a formulation
tailored to the transitions between full non-linear
and completely linear phases.
Our working premise will be that, during those linear-nonlinear (LNL) 
transitions, ``democracy" does not apply when determining the
framework needed to handle the gravitational fields.
That is, we expect that nonlinear effects are not equally
turned on and off for all the field variables.
There is evidence for such behavior.  
In the numerical study of neutron star binaries,
Wilson and collaborators \cite{Wilson95,Wilson96} used a simplified version of
the Einstein equations in which only a reduced set of metric variables
have nonlinear dynamics, with the remaining fields fixed in time. 
Although, a debate arose on one of the predictions from this work, namely
the collapse of the neutron stars before their coalescence
(see \cite{enna} for a resolution to this debate), this approach 
provides a reasonable approximation in the regime when the dynamics is dominated
by volume (conformal) deformations and the energy radiated per orbit is 
much smaller than the energy of orbital motion.
Our LNL formulation of Einstein equations builds upon Wilson's idea,
allowing for more general background space-times and the presence of
linear perturbations. The LNL 
is constructed by identifying, among the dynamical variables,
those quantities that can be kept fixed, those that can be treated perturbatively
and finally those that follow nonlinear dynamics. 
The hope is to arrive to a reduced system of Einstein equations that is not only 
more amenable to be handled numerically but also allows one to gain 
a better insight of the problem.

As customary in Numerical Relativity, the starting point is the 3+1 or ADM form of Einstein 
equations \cite{ADM62}. Under the 3+1 scheme, the space-time is viewed 
as the time history of a
foliation of space-like hypersurfaces with intrinsic metric $\hat h_{ab}$,
with Latin and Greek letters denoting spatial and space-time indices, respectively. 
Since most of our expressions will eventually involve conformal quantities,
we will reverse the usual convention and denote physical quantities with 
hats and conformal quantities without hats.
The 3+1 line element takes the form:
\begin{equation}
ds^2 = -\hat\alpha^2 dt^2 + \hat h_{ab}(dx^a 
+ \hat\beta^a dt)(dx^b + \hat\beta^b dt) \, ,
\label{eq:metric}
\end{equation}
with $\hat\alpha$ the lapse function characterizing proper time along the
normals to the hypersurface, and $\hat\beta^a$ the shift vector, representing
the freedom of relabeling spatial coordinate points in that surface. 
With the space-time metric given by (\ref{eq:metric}), Einstein equations 
take the form:
\begin{eqnarray}
\hat R + \hat K^2 - \hat K_{ab} \hat K^{ab}  & = & 0 
\label{eq:1.1}\\
\widehat \nabla_b \hat K^{ab}-\widehat\nabla^a \hat K & = & 0  
\label{eq:1.2}\\
\hat\partial_o \hat h_{ab} & = & -2\,\hat\alpha\,  \hat K_{ab} 
\label{eq:1.3}\\
\hat\partial_o \hat K_{ab} + 2\,\hat\alpha\,\hat K_{ac}\,\hat K^c\,_b 
- \hat\alpha\,\hat K_{ab}\, \hat K & = & 
\,\hat\alpha\,(\hat R_{ab}-\hat a_{ab}-\hat a_{a}\hat a_{b}) 
\label{eq:1.4}
\end{eqnarray}
Above, $\hat R_{ab}$ is the Ricci tensor constructed from the 3-metric $\hat h_{ab}$,
$\hat R = \hat R_a\,^a$ is the scalar curvature, and $\hat K_{ab}$ denotes the extrinsic curvature,
with $\hat K$ its trace. Also, $\hat a_a =  \widehat\nabla_a \ln{\hat \alpha}$
is the acceleration of Eulerian observers, and we introduce the tensor 
$\hat a_{ab} = \widehat\nabla_{a}\hat a_{b} = \hat a_{ba}$,
with $\hat a = \hat a_a\,^a = \widehat\nabla_{a}\hat a^{a}$. 
Covariant differentiation with respect to $\hat h_{ab}$ is
denoted by $\widehat \nabla_a$, 
and $\hat\partial_o \equiv \partial_t - \Lie_{\hat\beta}$.
The operator $\hat\alpha^{-1}\,\hat\partial_o$ is the derivative
with respect to proper time along the normal to the 
space-like hypersurfaces, and $\hat t^{\mu} = \hat\alpha\,\hat n^{\mu} + \hat\beta^{\mu}$ is
the time vector tangent to the congruence of world lines of coordinate observers.
Equations~(\ref{eq:1.1}) and (\ref{eq:1.2}) are respectively the Hamiltonian and
momentum constraints. Equation~(\ref{eq:1.3}), on the other hand, 
follows directly from the definition of the 
extrinsic curvature $\hat K_{ab}$.
For simplicity, we are only considering vacuum space-times; however, it is
straightforward to include matter sources.

The first step to derive the LNL formulation of Einstein equations involves 
splitting the evolution equations for the 
metric $\hat h_{ab}$ and extrinsic curvature $\hat K_{ab}$ into trace and trace-free
equations. That is, using $\hat K_{ab} = \hat A_{ab} + \hat h_{ab}\, \hat K /3$ with
$\hat A_a\,^a = 0$,
Eqs.~(\ref{eq:1.1}-\ref{eq:1.4}) are rewritten as:
\begin{eqnarray}
     \hat R + \frac{2}{3}\hat K^2 - \hat A_{ab} \hat A^{ab}  & = & 0 
\label{eq:2.1}\\
     \widehat \nabla_b\hat A^{ab}-\frac{2}{3}\widehat \nabla^a\hat K & = & 0  
\label{eq:2.2}\\
     \hat\partial_o \ln{\hat h^{1/2}} & = & 
-\hat\alpha\,\hat K 
\label{eq:2.3}\\
     \hat \partial_o \hat K - \hat\alpha\,\hat K^2 & = &  
         \hat\alpha\,(\hat R- \hat a - \hat a_a\hat a^a)  
\label{eq:2.5}\\
     \hat h^{1/3} \hat\partial_o (\hat h^{-1/3} \hat h_{ab}) & = & 
     - 2\,\hat \alpha\, \hat A_{ab} 
\label{eq:2.4}\\
     \hat \partial_o \hat A_{ab} 
+ 2\,\hat\alpha\,\hat A_{ac}\hat A^c\,_b - \frac{1}{3}\,\hat\alpha\,\hat A_{ab}\, \hat K & = & 
\,\hat\alpha\,\langle\hat R_{ab}- \hat a_{ab}-\hat a_{a}\hat a_{b}\rangle 
\label{eq:2.6}
\end{eqnarray}
where angle brackets denote traceless symmetrization,
i.e. $\langle T_{ab}\rangle \equiv T_{(ab)} - \frac{1}{3}\,h_{ab}\,T$  
with $T = T_a\,^a$.

In numerical relativity, conformal transformations have been extremely useful, in particular
in the construction of initial
data \cite{York79} to single out those ``pieces" within $\hat h_{ab}$
and $\hat A_{ab}$ that are fixed by the constraints (\ref{eq:2.1}) and (\ref{eq:2.2}).
Recently, conformal transformations have been considered also
in formulations of the evolution equations (\ref{eq:2.3}-\ref{eq:2.6}).
There are some indications that these formulations exhibit 
improvements in numerical stability properties \cite{Shibata,Baumgarte,Alcubierre}.
Here, we introduce conformal transformation as a tool for factoring out nonlinearities. 
There is, however, a certain degree of ambiguity regarding the particular form of these
transformation rules.
Once the conformal transformation rule for the spatial metric $\hat h_{ab}$ is fixed, there is
no specific recipe for choosing the remaining transformation rules for the extrinsic curvature 
$\hat K_{ab}$, lapse function $\hat\alpha$ and shift vector $\hat\beta^a$. 
For instance, in the construction of initial data, York's choices \cite{York79} were mostly aimed at 
simplifying the task of solving the constraints (\ref{eq:2.1}) and (\ref{eq:2.2}).
After several attempts, we found that the most convenient conformal transformations that 
yield a LNL factorization are:
\begin{eqnarray}
\hat h_{ab} &=& e^{4\,\Omega} h_{ab}
\label{eq:t.1}\\
\hat h^{1/2} &=& e^{6\,\Omega}
\label{eq:t.2}\\
\hat A_{ab} &=& e^{2\,\Omega}  A_{ab}
\label{eq:t.3} \\
\hat K &=& e^{-2\,\Omega} K
\label{eq:t.4} \\
\hat\alpha &=&  e^{2\,\Omega}\alpha
\label{eq:t.5}\\
\hat\beta^a &=& \beta^a \,.
\label{eq:t.6}
\end{eqnarray}
Conformal transformations similar to the above
were used by Geroch \cite{geroch} to study the structure of the gravitational field 
at infinity.
The conformal transformation (\ref{eq:t.1}) for the spatial metric is just the standard
rule used in most of the literature. 
The transformation (\ref{eq:t.2}) for the determinant of $\hat h_{ab}$
implies that $h=1$
\cite{Shibata,Baumgarte,Alcubierre}.
This choice is always possible and simplifies the form
of the equations while keeping the procedure general.
A consequence of $h=1$, together with the conformal transformations (\ref{eq:t.1}-\ref{eq:t.6}),
is that $h_{ab}$ is a tensor density of weight $-2/3$ and $A_{ab}$ a tensor density 
of weight $-1/3$. Similarly, $K$ is a scalar density of weight $1/3$ and $\alpha$ a scalar density
of weight $-1/3$. 
Notice also that, contrary to the common practice of not assigning a conformal transformation
to the trace of the extrinsic curvature $K$, we have chosen a conformal transformation
for $K$ such that $\hat K_{ab} = e^{2\,\Omega}  K_{ab}$. 
It is also important to point out that with the above conformal transformations,
the line element (\ref{eq:metric}) takes the form
\begin{equation}
ds^2 = e^{4\,\Omega}[-\alpha^2 dt^2 + h_{ab}(dx^a 
+ \beta^a dt)(dx^b + \beta^b dt)] \, .
\label{eq:cmetric}
\end{equation}
Therefore, the space-time metric has the same transformation rule as the spatial metric. 
Furthermore, the transformations (\ref{eq:t.1}-\ref{eq:t.6}) yield quantities
$\alpha$ and $\beta$ that are 
the lapse function and shift vectors of the conformal space-time, which would not 
necessarily be the case in general. 

Given (\ref{eq:t.1}-\ref{eq:t.6}), the following conformal relationships are also obtained:
\begin{eqnarray}
\hat R_{ab} &=& R_{ab} - 2\,\omega_{ab} + 4\,\omega_a\omega_b
-2\,h_{ab}(\omega+2\,\omega_c\,\omega^c) \label{eq:t.7}\\
\hat R & = & e^{-4\,\Omega}(R - 8\,\omega -8\,\omega_a\omega^a) \label{eq:t.8}\\
\hat a_a &=& a_a + 2\,\omega_a  \label{eq:t.9}\\
\hat a_{ab} &=& a_{ab} + 2\,\omega_{ab} - 4\,a_{(a}\,\omega_{b)}
- 8\,\omega_a\,\omega_b + 2\,h_{ab}(a_c\,\omega^c + 2\,\omega_c\,\omega^c) \label{eq:t.10}\\
\hat a &=& e^{-4\,\Omega}\,(a + 2\,\omega + 2\,a_a\,\omega^a 
+ 4\,\omega_a\,\omega^a)\,, \label{eq:t.11} 
\end{eqnarray}
where $\omega_a = \nabla_{a}\Omega$ and
$\omega_{ab} = \nabla_{a} \omega_{b} = \omega_{ba}$
with $\omega = \omega_a\,^a = \nabla_{a}\omega^{a}$.
We also have that 
$\hat\partial_o =\partial_o$ and $\langle \widehat\nabla^a\hat\beta^b\rangle = e^{-4\,\Omega}\,
\langle\nabla^a\beta^b\rangle$.
With the set of conformal transformations (\ref{eq:t.1}-\ref{eq:t.11}), 
Einstein equations~(\ref{eq:2.1}-\ref{eq:2.6}) 
can now be rewritten as:
\begin{eqnarray}
R+\frac{2}{3}\,K^2-A_{ab}A^{ab} & = & 8\,(\omega+\omega_a\,\omega^a)
\label{eq:4.1}\\
\nabla_b A^{ab} - \frac{2}{3}\,\nabla^aK &=& 
-4\,\omega_b\,(A^{ab}+\frac{1}{3}\,h^{ab}K)
\label{eq:4.2}\\
     \partial_o \Omega & = & -\frac{1}{6}\,\alpha K
\label{eq:4.3}\\
     \partial_o K - \frac{2}{3}\,\alpha\,K^2 & = &
     \alpha\,(R-a-a_a a^a - \Pi) 
\label{eq:4.4}\\
     \partial_o h_{ab} & = & - 2\, \alpha A_{ab}
\label{eq:4.5}\\
     \partial_o A_{ab}+2\,\alpha\,A_{ac}\,A^c\,_b-\frac{2}{3}\,\alpha\,A_{ab}\,K
     & = & \alpha\,\langle R_{ab}-a_{ab}-a_a\,a_b-\Pi_{ab}\rangle\,,
\label{eq:4.6}
\end{eqnarray}
where
\ba
\Pi_{ab} &=& 4\,\omega_{ab}
-8\,\omega_a\,\omega_b+ 2\,h_{ab}\,(\omega+4\,\omega_c\,\omega^c
+a_c\,\omega^c)\\
\Pi &=& h^{ab}\,\Pi_{ab} = 10\,\omega + 16\,\omega_a\,\omega^a + 6\,a_a\,\omega^a \,.
\ea

In the form given by Eqs.~(\ref{eq:4.3}-\ref{eq:4.6}), the Einstein evolutions equations are
split into two sectors: $(\Omega,\,K)$ and $(h_{ab},\,A_{ab})$.  
The $(\Omega,\,K)$ sector is kept fully nonlinear, and, as we shall see later, 
approximations are introduced only when handling
$(h_{ab},\,A_{ab})$, specifically in the transverse components of those fields.
We start by rewriting the evolution equation (\ref{eq:4.5}) of the conformal metric as
\be
\partial_t h_{ab} - \Lie_{\beta} h_{ab} = \partial_t h_{ab} - 2\,\langle \nabla_a\,\beta_b\rangle
=  - 2\,\alpha A_{ab}\,,
\label{eq:5}
\ee
where the substitution $\Lie_{\beta} h_{ab} = 2\,\langle \nabla_a\,\beta_b\rangle$
in the second equality uses the condition that $h_{ab}$ is a tensor density of weight $-2/3$.
Next, we decompose the tensor $\alpha\,A_{ab}$ into its transverse and longitudinal
parts, namely
\be
\alpha\,A_{ab} = T_{ab} + L_{ab}  \,,
\label{eq:tl}
\ee
such that $\nabla_b\,T^{ab} = T = 0$ and $L_{ab} = \langle \nabla_a\,\xi_b\rangle$.
Substitution of (\ref{eq:tl}) into (\ref{eq:5}) yields
\be
\partial_t h_{ab} = 2\,\langle \nabla_a\,(\beta_b-\xi_b)\rangle - 2\,T_{ab}\,.
\label{eq:xx}
\ee
Since our goal is to treat the transverse-traceless fields perturbatively, 
it is then natural to choose the shift vector $\beta^a$ such that it
eliminates the longitudinal terms in (\ref{eq:5}); 
that is, $\beta^a = \xi^a$. With this choice, 
Eq.~(\ref{eq:xx}) becomes
\ba
\partial_t h_{ab} &=& - 2\,T_{ab}
\label{eq:44.4}\\
L_{ab} &=& \langle \nabla_a\,\beta_b\rangle\,.
\label{eq:lab}
\ea
Notice that this choice of shift vector implies that the conformal space satisfies 
a gauge condition belonging to the class of ``radiation gauges" \cite{smarr-york}, namely
\be
\nabla_b\partial_t(h^N\, h^{ab}) = 0\,,
\label{eq:11}
\ee
with $h = 1$ and $N=1/3$.
Condition (\ref{eq:11}), however, does not translate to 
a radiation gauge condition in the physical space-time since
\be
\widehat\nabla_b\partial_t (\hat h^{1/3}\,\hat h^{ab}) 
= \nabla_b\partial_t h^{ab} + 10\,\omega_b\,\partial_t h^{ab}\,,
\ee
and in general the second term of this equation does not vanish.

From the decomposition (\ref{eq:tl}) and the transverseness of $T_{ab}$, it follows that
\be
\nabla_b L^{ab} = \nabla_b (\alpha\,A^{ab})\,,
\label{eq:gauge0}
\ee
which can be rewritten using Eq.~(\ref{eq:4.2}) as
\be
\nabla_b L^{ab} - (a_b-4\,\omega_b)\,(L^{ab}+T^{ab}) = 
\frac{2}{3}\, \alpha\,(\nabla^aK - 2\,\omega^a\,K)\,.
\label{eq:gauge}
\ee
This equation is simply the momentum constraint and must be satisfied independent of the choice
of shift vector. However, when the specific gauge choice (\ref{eq:lab}) is made, Eq.~(\ref{eq:gauge})
is alse viewed as a gauge condition from which the shift vector is constructed.
A more geometrical interpretation of the gauge condition (\ref{eq:gauge})
is obtained by noticing that this condition, also known as ``minimal strian 
gauge," can be derived from
a global minimization of the time rate of change of the metric in the
conformal space \cite{smarr-york}, that is, from the variational principle
\be
\delta S[\alpha,\,\beta^a] = \delta\int \partial_t h_{ab}\,
\partial_t h_{cd}\,h^{ac}\,h^{bd}\,dv = 0\,,
\label{eq:variational}
\ee
varying $\beta^a$. We must emphasize that
condition (\ref{eq:gauge0}), or equivalently
(\ref{eq:gauge}), is a minimal strain gauge not applied to
the physical space-time but to the conformal foliation. 

The freedom of imposing a gauge condition in
connection with the lapse function 
$\alpha$ still remains. For instance, if the choice is to have maximal
conformal slices, i.e. $K= 0$, Eq.~(\ref{eq:4.4}) becomes an
elliptic equation for $\alpha$, namely $a+a_a a^a = R-\Pi$. 
This condition also implies
from Eq.~(\ref{eq:4.3}) that $\partial_o \Omega =0$; that is, the
conformal factor $\Omega$ is constant along the normals to the 
hypersurfaces. 
Other possible choices are the use the generalized harmonic gauge, 
$a_o + \alpha\,K = \alpha\,f$,
with $f$ an arbitrary function, or a lapse function constructed from
the variational principle (\ref{eq:variational}) by varying $\alpha$
instead of $\beta^a$. The latter approach yields
\be
\alpha\,A_{ab}\,A^{ab} = \langle \nabla_a\,\beta_b\rangle\,A^{ab}\,,
\label{eq:agauge0}
\ee
which with the help of the Hamiltonian constraint (\ref{eq:4.1}) can be rewritten
as:
\be
\alpha = \frac{L_{ab}\,(T^{ab}+L^{ab})}
{R+\frac{2}{3}\,K^2 - 8\,(\omega+\omega_a\,\omega^a)}\,.
\label{eq:agauge}
\ee
Notice that with this choice of gauge, the equations for the lapse and shift
become a coupled set of equations \cite{Brady}. 

So far, no approximations have been made. The theory remains completely nonlinear.
The only condition is that the shift vector satisfies Eq.~(\ref{eq:gauge}), 
a condition introduced to
eliminate the longitudinal terms from the evolution
equation of the conformal metric.
We now assume that, in the physical systems
where our LNL form of Einstein equations apply, the non-linearities are dominated
by conformal (volume) deformations. 
As mentioned before, numerical simulations of neutron star binaries 
\cite{Wilson95,Wilson96,Shapiro,Bonazzola} have shown
that there is indeed a phase during the inspiral where this approximation holds.
The main difference in the present work from those studies is that we allow for 
the presence of perturbations entering via shear deformations
of a conformal background. That is, we
decompose the conformal metric as 
\be
h_{ab} = \gamma_{ab} + \Psi_{ab}\,,
\label{eq:metric_perturb}
\ee
with $\partial_t \gamma_{ab} = 0$ and
$\Psi_{ab}$ a perturbation. An immediate consequence of 
the assumption (\ref{eq:metric_perturb}) is that from Eq.~(\ref{eq:44.4})  
\be
\partial_t \Psi_{ab} = -2\,T_{ab}\,,
\label{eq:7}
\ee
so $T_{ab}$ becomes a perturbative quantity, in agreement with our goal 
of treating the ``radiative," transverse-traceless fields as perturbations.
Furthermore, having $\partial_t \gamma_{ab} = 0$ and
$T_{ab}$ transverse-traceless implies that Eq.~(\ref{eq:7}) yields 
\ba
\partial_t \Psi &=& 0 \label{eq:cond_1}\\
\partial_t \widetilde\nabla_b\Psi^{ab} &=& 0 \label{eq:cond_2}\,,
\ea
where $\Psi \equiv \gamma^{ab}\Psi_{ab}$. Above and from now on, tensors
have their indices 
lowered and raised with the background metric $\gamma_{ab}$,
and tildes denote quantities in connection with this metric.
A direct implication of Eqs.~(\ref{eq:cond_1}) and (\ref{eq:cond_2}) is that,
if we choose initially the perturbation of the metric
to be transverse and traceless, these conditions are preserved by 
the evolution equation (\ref{eq:7}). We will take advantage of
this property and assume that initially
$\widetilde\nabla_b\Psi^{ab} = \Psi = 0$ since this choice simplifies the equations
significantly. 

At this point, we have the Hamiltonian constraint (\ref{eq:4.1}), 
the gauge condition (\ref{eq:gauge}) on the shift vector and
the nonlinear evolution equations (\ref{eq:4.3}) and (\ref{eq:4.4})
for the conformal factor and the trace of the extrinsic curvature.
In addition, we have the evolution equation (\ref{eq:7}) for the perturbation to the
conformal metric; therefore, what remains 
is to obtain an evolution equation for $T_{ab}$. 

From Eq.~(\ref{eq:7}), we have that
\be
\partial_o \Psi_{ab} = -2\,\left(T_{ab} + \frac{1}{2}\,\Lie_\beta \Psi_{ab}\right)
         = -2\,{\cal T}_{ab}\,,
\label{eq:8}
\ee
where we have defined 
\be
{\cal T}_{ab} \equiv T_{ab} + \frac{1}{2}\,\Lie_\beta \Psi_{ab}\,.
\label{eq:tau}
\ee
On the other hand, Eq.~(\ref{eq:tl}) can be rewritten as:
\ba
\alpha\,A_{ab} &=& L_{ab} + T_{ab} = \langle \nabla_a\,\beta_b\rangle + T_{ab}\nonumber \\
    &=& \frac{1}{2}\,\Lie_\beta h_{ab} + T_{ab}
    = \frac{1}{2}\,\Lie_\beta \gamma_{ab} + \frac{1}{2}\,\Lie_\beta \Psi_{ab} + T_{ab}\nonumber \\
    &=& \langle\widetilde \nabla_a\,\beta_b\rangle+ {\cal T}_{ab} = \tilde L_{ab} + {\cal T}_{ab} \,,
\label{eq:10}
\ea
where we have used that $\gamma_{ab}$ is a tensor density of weight $-2/3$.
Substitution of (\ref{eq:10}) into (\ref{eq:4.6}) then yields
\ba
&&\partial_o (\alpha\,A_{ab}) 
- (a_o-4\,\partial_o\Omega)\,(\alpha\,A_{ab}) +2\,(\alpha\,A_{ac})(\alpha\,A^c\,_b) =\nonumber\\
&&\partial_o \tilde L_{ab} - (a_o-4\,\partial_o\Omega)\,\tilde L_{ab} 
+ 2\,\tilde L_{ac}\tilde L^c\,_b + \nonumber \\
&&\partial_o {\cal T}_{ab} - (a_o-4\,\partial_o\Omega)\,{\cal T}_{ab}   
+ 2\,(\tilde L_{ac}\, {\cal T}^c\,_b + \tilde L_{bc}\,{\cal T}^c\,_a)
- 2\,\tilde L_{ac}\tilde L_{db}\,\Psi^{cd} = \nonumber\\
&&\alpha^2 \,\langle R_{ab}-a_{ab}-a_a\,a_b-\Pi_{ab}\rangle\,,
\label{eq:lhs}
\ea
where 
\ba
\langle a_{ab} \rangle &=& \langle \tilde a_{ab} \rangle 
+\frac{1}{3}(\gamma_{ab}\Psi^{cd}-\gamma^{cd}\Psi_{ab})\tilde a_{cd}
- S^c\,_{ab}\,\tilde a_c 
\label{eq:12.2} \\
\langle a_a a_b \rangle &=& \langle \tilde a_a \tilde a_b \rangle
+\frac{1}{3}(\gamma_{ab}\Psi^{cd}-\gamma^{cd}\Psi_{ab})\tilde a_c\tilde a_d
\label{eq:12.3} \\
\langle R_{ab} \rangle &=& \langle \tilde R_{ab} \rangle 
+\frac{1}{3}(\gamma_{ab}\Psi^{cd}-\gamma^{cd}\Psi_{ab})\tilde R_{cd}
-\frac{1}{2} \widetilde\nabla_c\widetilde\nabla^c \Psi_{ab} + \tilde R_{c(a}\Psi_{b)}\,^c
+ \tilde R_{cabd}\Psi^{cd} 
\label{eq:12.1}\\
\langle \Pi_{ab} \rangle &=& \langle \tilde\Pi_{ab} \rangle 
+\frac{1}{3}(\gamma_{ab}\Psi^{cd}-\gamma^{cd}\Psi_{ab})\tilde \Pi_{cd}
-4\,S^c\,_{ab}\,\tilde \omega_c
+2\,\Psi_{ab}(\tilde\omega+4\,\tilde\omega_c
\tilde\omega^c + \tilde a_c\tilde\omega^c)\,,
\label{eq:12.4} 
\ea
with
\be
S^a\,_{bc} = \frac{1}{2}(\widetilde\nabla_b\,\Psi^a\,_c
+\widetilde\nabla_c\,\Psi^a\,_b-\widetilde\nabla^a\,\Psi_{bc}) \,;
\label{eq:13.1}
\ee
therefore, to zero-order Eq.~(\ref{eq:lhs}) reads
\be
\partial_o \tilde L_{ab} - (a_o-4\,\partial_o\Omega)\,\tilde L_{ab} + 2\,\tilde L_{ac}\tilde L^c\,_b =
    \alpha^2\,\langle \tilde R_{ab} - 
\tilde a_{ab}-\tilde a_a\,\tilde a_b-\tilde \Pi_{ab}\rangle\,.
\label{eq:14.1}
\ee
These evolution equations are redundant because, as mentioned before, 
the gauge conditions (\ref{eq:gauge}) that determine $\tilde L_{ab}$
are also the momentum contraints, which in turn preserve
Eqs.~(\ref{eq:14.1}).
To first-order in the perturbations
Eq.~(\ref{eq:lhs}) yields the evolution equation for the perturbation
${\cal T}_{ab}$:
\ba
&&\partial_o {\cal T}_{ab} - (a_o-4\,\partial_o\Omega)\,{\cal T}_{ab} 
+ 2\,(\tilde L_{ac}\, {\cal T}^c\,_b + \tilde L_{bc}\,{\cal T}^c\,_a)
- 2\,\tilde L_{ac}\tilde L_{db}\,\Psi^{cd} = \nonumber \\
&& \alpha^2\, \Biggl\lbrack
-\frac{1}{2} \widetilde\nabla_c\widetilde\nabla^c  \Psi_{ab} + \tilde R_{c(a}\Psi_{b)}\,^c
+ \tilde R_{cabd}\Psi^{cd} 
+ S^c\,_{ab}(\tilde a_c+4\,\tilde\omega_c) 
- 2\,\Psi_{ab}(\tilde\omega+4\,\tilde\omega_c
\,\tilde\omega^c + \tilde a_c\,\tilde\omega^c) \nonumber\\
&&+ \frac{1}{3} (\gamma_{ab}\Psi^{cd}-\gamma^{cd}\Psi_{ab})(
\tilde R_{cd} - \tilde a_{cd}-\tilde a_c\,\tilde a_d-\tilde \Pi_{cd})\Biggr\rbrack\,.
\label{eq:14.2}
\ea
By combining this equation with Eq.~(\ref{eq:8}), one obtains 
\be
\left(-\frac{1}{\alpha}\partial_o \frac{1}{\alpha}\partial_o 
+ \widetilde\nabla_c\widetilde\nabla^c\right) \Psi_{ab} = J_{ab}\,,
\label{eq:15.a}
\ee
where 
\ba
\frac{1}{2}\alpha^2\,J_{ab} &=& \frac{2}{3}\,\alpha\,K\,{\cal T}_{ab}
- 2\,(\tilde L_{ac}\, {\cal T}^c\,_b + \tilde L_{bc}\,{\cal T}^c\,_a)
+ 2\,\tilde L_{ac}\tilde L_{db}\,\Psi^{cd} \nonumber \\
&+& \alpha^2 \Biggl\lbrack
\tilde R_{c(a}\Psi_{b)}\,^c 
+ \tilde R_{cabd}\Psi^{cd}
+ S^c\,_{ab}(\tilde a_c+4\,\tilde\omega_c)
- 2\,\Psi_{ab}(\tilde\omega+4\,\tilde\omega_c
\,\tilde\omega^c + \tilde a_c\,\tilde\omega^c) \nonumber\\
&+& \frac{1}{3} (\gamma_{ab}\Psi^{cd}-\gamma^{cd}\Psi_{ab})(
\tilde R_{cd} - \tilde a_{cd}-\tilde a_c\,\tilde a_d-\tilde \Pi_{cd})\Biggr\rbrack\,.
\label{eq:15.b}
\ea
That is, the perturbation $\Psi_{ab}$ obeys a wave equation in the conformal space-time.

One more simplification of the LNL form of Einstein's equations is possible:
neglecting the 
back-reaction of the perturbative fields ($\Psi_{ab},{\cal T}_{ab}$) on the
nonlinear ($\Omega,K$) and gauge ($\alpha,\beta^a$) fields.
Under this approximation, covariant differentiation and the raising and lowering of indices
are performed with
the background metric $\gamma_{ab}$.  

In summary, the following picture emerges. 
Starting from the standard 3+1
formulation of Einstein equations in terms of variables 
$(\hat h_{ab},\hat K_{ab})$, we have constructed a formulation
in which the field variables are divided into:
\begin{enumerate}
\item Nonlinear fields $(\Omega, K)$ with their dynamics given by
\ba
\partial_o\, \Omega & = & -\frac{1}{6}\,\alpha K
\label{eq:9.1}\\
\partial_o\, K -\frac{2}{3}\,\alpha\,K^2 &=& \alpha\,
(\tilde R-\tilde a-\tilde a_a \tilde a^a - \tilde \Pi) \,.
\label{eq:9.2}
\ea
\item Perturbative fields $(\Psi_{ab},{\cal T}_{ab})$ with equations of motion
\ba
\partial_o \Psi_{ab} &=& -2\,{\cal T}_{ab}
\label{eq:9.3}\\
\partial_o {\cal T}_{ab} - a_o\,{\cal T}_{ab} &=&
-\frac{1}{2} \alpha^2(\widetilde\nabla_c\widetilde\nabla^c \Psi_{ab} - J_{ab})\,.
\label{eq:9.4}
\ea
\item A background conformal metric $\gamma_{ab}$ such that 
$\partial_t \gamma_{ab} = 0$.
\item And a shift vector $\beta^a$ satisfying the condition
\be
\widetilde\nabla_b \tilde L^{ab} - (\tilde a_b-4\,\tilde\omega_b)\,\tilde L^{ab} =
\frac{2}{3}\, \alpha\,(\widetilde\nabla^aK - 2\,\tilde\omega^a\,K)\,.
\label{eq:9.5}
\ee
\end{enumerate}
The conditions and assumptions used to 
derive this LNL system of equations are:
\begin{enumerate}
\item The field variables are conformally transformed according to (\ref{eq:t.1}-\ref{eq:t.6}).
\item The shift vector is chosen to provide the longitudinal part of $\alpha\,A_{ab}$,
which in turns implies that it must satisfy Eq.~(\ref{eq:9.5}).
\item The conformal metric $h_{ab}$ is decomposed into a background metric $\gamma_{ab}$ plus
a transverse-traceless perturbation $\Psi_{ab}$.
\item The back-reaction from the perturbations is ignored. 
\end{enumerate}

As mentioned before, the only gauge freedom remaining 
is that in connection with the lapse 
function $\alpha$. Furthermore,
if a fully constrained evolution is preferred, the conformal factor could be
obtained from the Hamiltonian constraint (\ref{eq:4.1}).
When in addition one assumes that the metric perturbation $\Psi_{ab}$,
and consequently ${\cal T}_{ab}$, could be neglected, the LNL system reduces to
a nonlinear system similar to that used by Wilson and collaborators
\cite{Wilson95,Wilson96} for studying neutron star collisions.
The differences would be due to our particular choice of conformal transformations.

In conclusion, with the help of suitable conformal transformations 
and a shift vector gauge condition,
we have constructed an approximate formulation of
Einstein equations that could yield advantages in the study of gravitational systems 
where radiative, transverse-traceless variables can be treated as perturbations 
of a fixed, background spatial metric,
with the remaining fields possessing nonlinear dynamics.
Since this separation depends on the choice  of conformal transformations,
it remains to be tested whether the conformal transformations used in our derivation
of the LNL system of equations indeed constitute an appropriate approximation.
When the LNL system is applied to the {\it Inspiral-Merging} intermediate regime,
an interesting approach suggested in Ref.~\cite{Brady} that could place 
our approximation in a more firmer grounds, is to perform the calculations
in a co-rotating frame. The objective under this approach 
is to find a spatial vector $b^{\mu}$, i.e. $b_{\mu}\,n^{\mu}= 0$ with 
$n^{\mu}$ the normal to the constant $t$ hypersurfaces, such that 
$\xi^{\mu} = \alpha\,n^{\mu} + \beta^{\mu} + b^{\mu}$
is an ``almost Killing vector field," namely
$\Lie_\xi g_{\mu\nu} \approx 0$.

Special thanks to Abhay Ashtekar, Mijan Huq, Richard Matzner, Philip Papadopoulos
and Deirdre Shoemaker
for helpful discussions. 
This work was supported by NSF grants PHY 98-00973 and PHY 93-57219 (NYI).


\begin{references}

\def\prl#1#2#3{{ Phys. Rev. Lett.\ }, {\bf #1}, #2 (#3)}
\def\prd#1#2#3{{ Phys. Rev. D}, {\bf #1}, #2 (#3)}
\def\plb#1#2#3{{ Phys. Lett. B}, {\bf #1}, #2 (#3)}
\def\prep#1#2#3{{ Phys. Reports}, {\bf #1}, #2 (#3)}
\def\phys#1#2#3{{ Physica}, {\bf #1}, #2 (#3)}
\def\jcp#1#2#3{{ J. Comput. Phys.}, {\bf #1}, #2 (#3)}
\def\jmp#1#2#3{{ J. Math. Phys.}, {\bf #1}, #2 (#3)}
\def\cpr#1#2#3{{ Computer Phys. Rept.}, {\bf #1}, #2 (#3)}
\def\cqg#1#2#3{{ Class. Quantum Grav.}, {\bf #1}, #2 (#3)}
\def\cma#1#2#3{{ Computers Math. Applic.}, {\bf #1}, #2 (#3)}
\def\mc#1#2#3{{ Math. Compt.}, {\bf #1}, #2 (#3)}
\def\apj#1#2#3{{ Astrophys. J.}, {\bf #1}, #2 (#3)}
\def\apjs#1#2#3{{ Astrophys. J. Suppl.}, {\bf #1}, #2 (#3)}
\def\acta#1#2#3{{ Acta Astronomica}, {\bf #1}, #2 (#3)}
\def\sa#1#2#3{{ Sov. Astro.}, {\bf #1}, #2 (#3)}
\def\sia#1#2#3{{ SIAM J. Sci. Statist. Comput.}, {\bf #1}, #2 (#3)}
\def\aa#1#2#3{{ Astron. Astrophys.}, {\bf #1}, #2 (#3)}
\def\mnras#1#2#3{{ Mon. Not. R. astr. Soc.}, {\bf #1}, #2 (#3)}
\def\prsla#1#2#3{{ Proc. R. Soc. London, Ser. A}, {\bf #1}, #2 (#3)}
\def\ijmpc#1#2#3{{ I.J.M.P.} C {\bf #1}, #2 (#3)}

\bibitem{price-pullin} R.H. Price and J. Pullin
\prl{72}{3297}{1994}

\bibitem{Alliance} For current status of the Binary Black Hole Grand Challenge,
see http://www.npac.syr.edy/projects/bh

\bibitem{Brady} P.R. Brady, J.D.E. Creighton and K.S. Thorne
\prd{58}{061501}{1998}

\bibitem{Wilson95} J.R.~Wilson and G.J.~Mathews,
\prl{75}{4161}{1995}

\bibitem{Wilson96} J.R.~Wilson, G.J.~Mathews and P.~Marronetti,
\prd{54}{1317}{1996}

\bibitem{enna} E.E.~Flanagan, \prl{82}{1354}{1999}

\bibitem{ADM62} R.~Arnowitt, S.~Deser and C.W.~Misner:
in {\sl Gravitation} ed. L.~Witten (Wiley, New York, 1962) p. 227.

\bibitem{York79} J.W.~York, Jr., in 
{sl Sources of Gravitational Radiation}, ed. L. Smarr 
(Cambridge University Press, Cambridge, England, 1979), p. 83.

\bibitem{Shibata} M. Shibata and T. Nakamura, \prd{52}{5428}{1995}

\bibitem{Baumgarte} T.W. Baumgarte and S.L. Shapiro,
\prd{59}{024007}{1999}

\bibitem{Alcubierre} M. Alcubierre, B. Br\"ugmann, M. Miller and W.-M. Suen
gr-qc/9903030

\bibitem{geroch} R. Geroch \jmp{13}{956}{1972}

\bibitem{smarr-york} L. Smarr and J.W.~York, Jr.,
\prd{17}{1945}{1978}

\bibitem{Shapiro} T.W. Baumgarte, G.B. Cook, M.A. Scheel, S.L. Shapiro 
and S.A. Teukolsky, \prl{79}{1182}{1997}

\bibitem{Bonazzola} S. Bonazzola, E. Gourgoulhon and J.-A. Marck, 
\prl{82}{892}{1999}

\end{references}
\end{document}